\documentclass[nofootinbib,showpacs]{revtex4}

\usepackage{graphicx}

\newcommand{\be}{\begin{equation}}
\newcommand{\ee}{\end{equation}}
\newcommand{\ba}{\begin{eqnarray}}
\newcommand{\ea}{\end{eqnarray}}
\newcommand{\ban}{\begin{eqnarray*}}
\newcommand{\ean}{\end{eqnarray*}}

\begin{document}

\title{EVOLUTION OF ANISOTROPY OF PARTON SYSTEM FROM \\
RELATIVISTIC HEAVY-ION COLLISIONS}

\author{Weronika Jas\footnote{Electronic addres:
{\tt weronika.jas@gmail.com}}}

\affiliation{Institute of Physics, \'Swi\c etokrzyska Academy \\
ul.~\'Swi\c etokrzyska 15, PL - 25-406 Kielce, Poland}

\author{Stanis\l aw Mr\' owczy\' nski\footnote{Electronic address:
{\tt mrow@fuw.edu.pl}}}

\affiliation{ Institute of Physics, \'Swi\c etokrzyska Academy \\
ul.~\'Swi\c etokrzyska 15, PL - 25-406 Kielce, Poland \\
and So\l tan Institute for Nuclear Studies \\
ul.~Ho\.za 69, PL - 00-681 Warsaw, Poland}

\date{3-rd September 2007}

\begin{abstract}

Evolution of anisotropy in momentum and coordinate space
of the parton system produced in relativistic heavy-ion
collisions is discussed within the free-streaming 
approximation. The momentum distribution evolves from
the prolate shape - elongated along the beam - to the 
oblate one - squeezed along the beam. At the same time 
the eccentricity in the coordinate space, which occurs 
at finite values of impact parameter, decreases. It is 
argued that the parton system reaches local thermodynamic 
equilibrium before the momentum distribution becomes oblate. 
 
\end{abstract}



\maketitle


\section{Introduction}


Parton system, which emerges at the early stage of relativistic
heavy-ion collisions, is anisotropic both in momentum and coordinate
space. The anisotropies crucially influence the dynamics of the system:
the momentum one causes plasma color instabilities (for a review see 
\cite{Mrowczynski:2005ki}); the coordinate-space one is responsible 
for the hydrodynamic elliptic flow (for a review see \cite{Kolb:2003dz}).
An eccentricity of the overlap region of colliding nuclei at nonzero 
impact parameter decreases when the parton system produced in the 
overlap region expands. While the eccentricity simply decays, the 
parton momentum distribution, which is observed locally, changes from 
a strongly prolate shape - elongated along the beam axis - to an oblate 
form - squeezed along the beam. 

Since hydrodynamics requires at least partial local equilibrium
\cite{Arnold:2004ti}, an observation of the elliptic flow suggests 
that the parton system is thermalized before the initial eccentricity 
is significantly reduced. The equilibration time of the parton system 
$t_{\rm eq}$ was actually estimated to be shorter than 1 fm/{\it c} 
\cite{Heinz:2004pj}. Inter-parton collisions cannot equilibrate the 
system so fast but magnetic unstable modes due to the momentum 
anisotropy speed-up the equilibration process. However, the
question arises whether the momentum distribution is prolate
or oblate just before the equilibrium is reached. In this note 
we attempt to resolve the issue in a very simple classical model 
where partons produced in the overlapping region of colliding nuclei 
freely escape from it. We analyze how the coordinate space anisotropy 
decays and how the momentum distribution evolves in a box which includes 
the Lorentz contracted region where the partons are initially produced. 
The effect of finite formation time of produced partons is taken into
account.

We are fully aware how naive our approach is. Matter, which emerges 
at the early stage of relativistic heavy-ion collisions, is very
dense, presumably inhomogeneous, and partially coherent due to
the memory of pure quantum state of two colliding nuclei. Such 
a system cannot be reliably described in terms of kinetic theory 
with weakly interacting quasi-particles on mass-shell. A derivation
of transport equation from quantum field theory clearly reveals
the limitations of the kinetic approach \cite{Mrowczynski:1989bu}.
We believe, however, that our free-streaming model still grasps
global features of the early stage system. The evolution of 
anisotropies, which is of our main interest here, is dominated 
by the system's expansion. And it proceeds with the velocity of
light independently of details of the system's dynamics. However,
it should be clearly understood that in the free-streaming model,
where no interaction is present, a real equilibration does not 
take place. Therefore, even if the particle momentum distribution 
appears to be of the equilibrium form, one cannot conclude that
the system has reached the equilibrium, which is required by the 
hydrodynamic description, as the pressure resulting from 
inter-particle collisions is absent in the free-streaming model.
We return to the discussion in the concluding section of the paper.


\section{Decay of eccentricity}
\label{sec-x-asym}

The partons are assumed to be produced in an ellipsoidal region 
parameterized by the three-dimensional Gaussian function centered
at zero with the widths: $\sigma_x$, $\sigma_y$, $\sigma_z$, 
where $x$, $y$ and $z$ denote Cartesian coordinates. As usually, 
the axis $z$ is along the beam. The distribution of parton's rapidity, 
which is denoted here by $Y$, is assumed to be Gaussian as well with
the width $\Delta\!Y$. We note that the rapidity distribution of
charged pions produced in Au-Au collisions at $\sqrt{s} = 200$ GeV 
per nucleon-nucleon pair is well described by the Gaussian distribution
with $\Delta\!Y =2.3$ \cite{Bearden:2004yx}. Since we work in the 
center-of-mass frame, the rapidity distribution is centered at zero. 
We also assume that partons are massless and then, as  it will be 
evident later on, we do not need to specify the distribution 
of their transverse momenta which is denoted as $P(p_T)$. Thus, the 
distribution function of partons, which obeys the collisionless
Boltzmann equation, is
\be
\label{dis-fun}
f(t,{\bf r}, {\bf p}) = 
\frac{1}
{\sigma_x \sigma_y \sigma_z \: \Delta\!Y \langle p_T \rangle}
\exp \bigg[- \frac{(x - v_xt)^2}{2 \sigma_x^2}
- \frac{(y - v_yt)^2}{2 \sigma_y^2}
- \frac{(z - v_zt)^2}{2 \sigma_z^2}
- \frac{Y^2}{2 \Delta\!Y^2 }\bigg] \; 
\frac{P(p_T)}{p_T^2 {\rm ch} Y} \;,
\ee
where 
$$
\langle p_T \rangle \equiv \int_0^\infty dp_T p_T P(p_T)
\;,\;\;\;\;\;
\int_0^\infty dp_T P(p_T) =1 \;,
$$
and the velocities $v_x$, $v_y$ and $v_z$ are given as
$$
v_x = \frac{\cos \phi}{{\rm ch} Y} \;,\;\;\;
v_y = \frac{\sin \phi}{{\rm ch} Y} \;,\;\;\;
v_z = {\rm th}Y \;,
$$
with $\phi$ being the azimuthal angle in the momentum space. 
We note that we use the natural units where $c = 1$. 
The distribution function is normalized to unity 
$$
\int d^3 r \frac{d^3 p}{(2\pi)^3} \, f(t,{\bf r}, {\bf p}) = 1\;.
$$

\begin{figure}[t]
\begin{minipage}{82mm}
\vspace{7mm}
\includegraphics*[width=82mm]{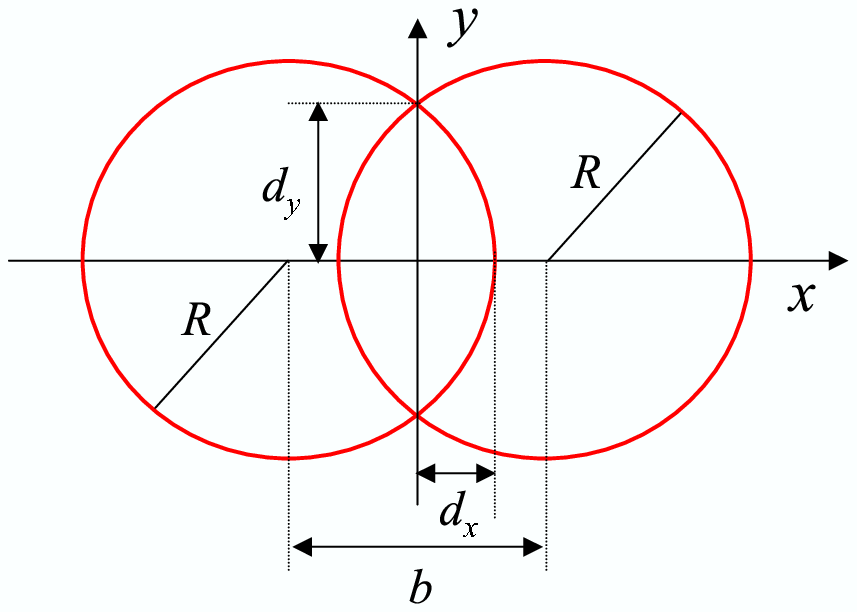}
\caption{(Color online) A view of colliding nuclei of equal radius 
$R$, as seen in the plane $x\!-\!y$ transverse to the beam axis.
\label{fig-geo}} 
\end{minipage}\hspace{4mm}
\begin{minipage}{85mm}
\vspace{0.7cm}
\includegraphics*[width=85mm]{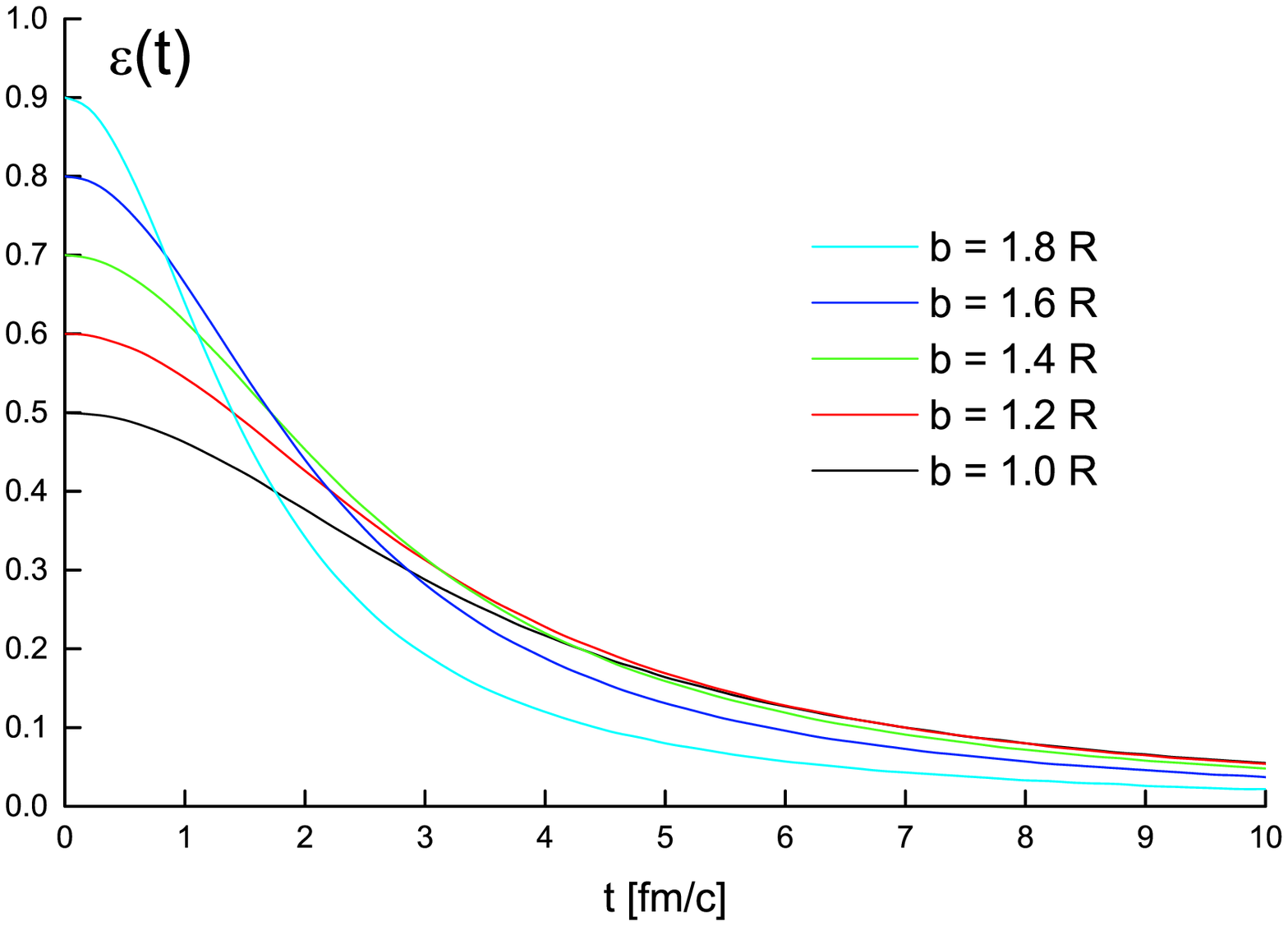}
\caption{(Color online) The eccentricity in Au-Au collisions as 
a function of time for five values of the impact parameter. The 
curve with the largest initial eccentricity corresponds to $b = 1.8 R$, 
the next one to $b = 1.6 R$, etc. 
\label{fig-eccen-x}} 
\end{minipage}
\end{figure}

When the impact parameter of the colliding nuclei is chosen to be
along the axis $x$, as shown in Fig.~\ref{fig-geo}, the eccentricity, 
which drives the elliptic flow, is defined as
\be
\label{eccen-x-def}
\varepsilon = \frac{\langle y^2 \rangle - \langle x^2 \rangle}
{\langle y^2 \rangle + \langle x^2 \rangle} \;.
\ee
With the distribution function (\ref{dis-fun}), one computes
$$
\langle x^2 \rangle \equiv \int d^3 r \frac{d^3 p}{(2\pi)^3} \, 
x^2 f(t,{\bf r}, {\bf p}) = 
\sigma_x^2 + \alpha \, t^2
$$
where 
$$
\alpha = 
\frac{1}
{2\sqrt{2\pi}\Delta\!Y}
\int_{-\infty}^{+\infty} \frac{dY}{{\rm ch}^2 Y}\,
\exp \bigg[- \frac{Y^2}{2 \Delta\!Y^2 }\bigg] \;.
$$
If one takes into account only partons with the rapidities $Y$
obeying $Y_{\rm min} < Y < Y_{\rm max}$, the formula, which defines
the coefficient $\alpha$, changes to 
$$
\alpha = 
\frac{\int_{Y_{\rm min}}^{Y_{\rm max}} dY {\rm ch}^{-2}Y \,
\exp \big[- \frac{Y^2}{2 \Delta\!Y^2 }\big] }
{2 \int_{Y_{\rm min}}^{Y_{\rm max}} dY \, 
\exp \big[- \frac{Y^2}{2 \Delta\!Y^2 }\big] } \;.
$$

Computing $\langle y^2 \rangle$ analogously to $\langle x^2 \rangle$,
one finds the eccentricity (\ref{eccen-x-def}) as
\be
\label{eccen-x}
\frac{\varepsilon (t)}{\varepsilon (0)} = 
\Big({1 + \frac{\alpha}{R_T^2} \, t^2}\Big)^{-1} \;,
\ee 
where the initial eccentricity $\varepsilon(0)$ and the average 
transverse size of the overlap region of colliding nuclei $R_T$ are
$$
\varepsilon(0) = \frac{\sigma_y^2 - \sigma_x^2}{\sigma_y^2 + \sigma_x^2}
\;,\;\;\;\;\;\;
R_T^2 \equiv \frac{\sigma_y^2 + \sigma_x^2}{2} \;.
$$ 
The formula (\ref{eccen-x}) was earlier derived in 
\cite{Kolb:2003dz,Kolb:2000sd} for a narrow interval around $Y=0$
when $\alpha = 1/2$. Unfortunately, by mistake $\alpha = 1$ in 
\cite{Kolb:2000sd}.

When the impact parameter varies, both the initial eccentricity  
$\varepsilon (0)$ and the average transverse size $R_T$ change. 
To express the two quantities through the nuclear radius $R$
(a hard sphere parameterization is adopted here) and the impact 
parameter $b$, we replace the overlap region of two circles shown 
in Fig.~\ref{fig-geo} by the ellipse given by the equation
$$
\frac{x^2}{d_x^2} + \frac{y^2}{d_y^2} \le 1
$$
with the half-axes $d_x$ and $d_y$ defined in Fig.~\ref{fig-geo}.
Elementary geometric arguments provide $d_x = R - b/2$ and
$d_y = \sqrt{R^2 - b^2/4}$. Since the mean square sizes of the 
ellipse are $\langle x^2 \rangle = d_x^2/4$ and 
$\langle y^2 \rangle = d_y^2/4$, we identify $\sigma_x$ and 
$\sigma_y$ with $d_x/2$ and $d_y/2$, respectively. Then, one 
expresses $\varepsilon (0)$ and $R_T$ through $R$ and $b$ as 
$$
\varepsilon(0) = \frac{b}{2R} \;,\;\;\;\;\;
R_T^2 = \frac{R^2}{4}(1 - \varepsilon (0)) \;.
$$

In Fig.~\ref{fig-eccen-x} we show predictions of the formula 
(\ref{eccen-x}) for Au-Au collisions at midrapidity ($\alpha=1/2$). 
The radius of a gold nucleus is chosen to be 7 fm. As seen in 
Fig.~\ref{fig-eccen-x}, the larger initial eccentricity, the 
faster its decay. We note that the largest elliptic flow in Au-Au 
collisions is observed at $b \approx 10 \; {\rm fm}$ \cite{Adler:2002pu} 
corresponding to $b \approx 1.4 R$. At larger impact parameters, the
produced system is presumably too small to fully manifests collective
hydrodynamic behavior. An analysis of the experimental elliptic flow 
data within the hydrodynamic model shows that the eccentricity 
cannot be reduced to more than 75\% of its  initial value 
\cite{Heinz:2004pj}, see also \cite{Drescher:2007cd}. When the 
reduction is larger, the ideal hydrodynamics, which gives an upper 
limit of the flow, significantly underestimates experimental data. 
Fig.~\ref{fig-eccen-x} shows that for $b \approx 1.4 R$ the eccentricity 
is reduced to 75\% of its initial value at 1.5 fm/$c$. Within this time 
interval the system has to be equilibrated to start hydrodynamic 
evolution responsible for the elliptic flow. Thus, our estimate of
the upper limit of equilibration time is 1.5 fm/$c$. Actually, the 
hydrodynamic analysis \cite{Heinz:2004pj}, which uses not only the 
elliptic flow data but other experimental constraints as well, 
provides the equilibration time $t_{\rm eq}$ as short as 0.6 fm/$c$.


\section{Momentum anisotropy evolution}
\label{sec-p-asym}

In this section we compute the momentum distribution of partons 
in a box of sizes $L_x$, $L_y$ and $L_z$. The box is centered at 
$x=y=z=0$. To simplify the calculations, the sharp edge box is 
replaced by the `box' function of Gaussian form
\be
\label{box-fun}
{\cal O}({\bf r}) = {\cal O}_x(x){\cal O}_y(y){\cal O}_z(z)
\;,\;\;\;
{\cal O}_i(r_i) = \sqrt{\frac{6}{\pi}} \; 
\exp \Big[- \frac{6r_i^2}{L_i^2}\Big], \; i=x,y,z ,
\ee
which obeys the conditions
$$
\int_{-L_i/2}^{L_i/2} dr_i = \int dr_i {\cal O}_i(r_i)
\;,\;\;\;\;\;\;\;
\int_{-L_i/2}^{L_i/2} dr_i \, r_i^2
=  \int dr_i \,r_i^2 {\cal O}_i(r_i) \;.
$$

\begin{figure}[t]
\begin{minipage}{83mm}
\includegraphics*[width=83mm]{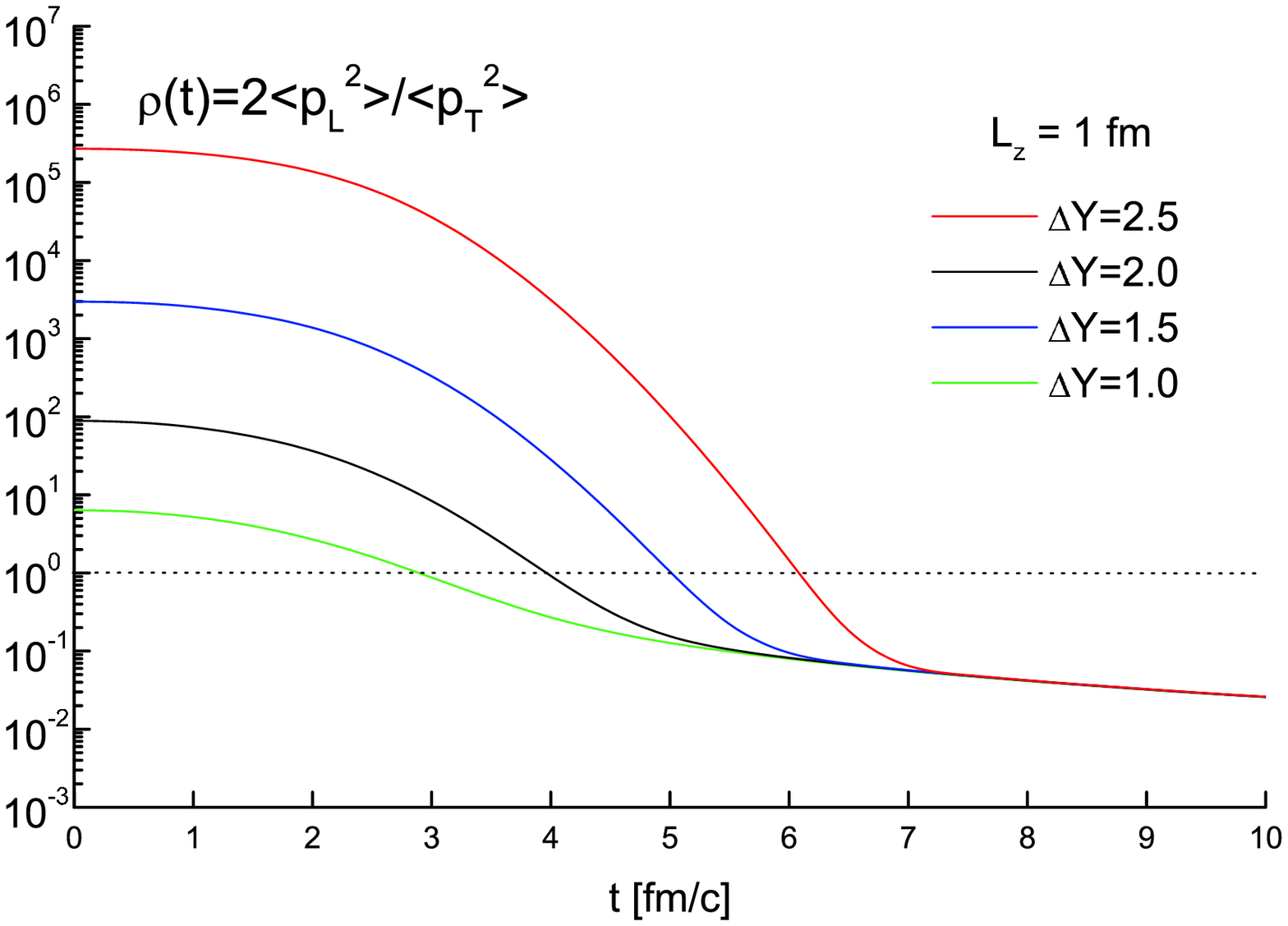}
\caption{(Color online) The momentum anisotropy as a function of 
time for four values of the rapidity distribution width $\Delta\!Y$.
The most upper line corresponds to $\Delta\!Y=2.5$, the lower one 
to $\Delta\!Y=2.0$, etc. The longitudinal size of the box $L_z$ 
equals 1 fm.
\label{fig-asym-p1}} 
\end{minipage}\hspace{4mm}
\begin{minipage}{83mm}
\includegraphics[width=83mm]{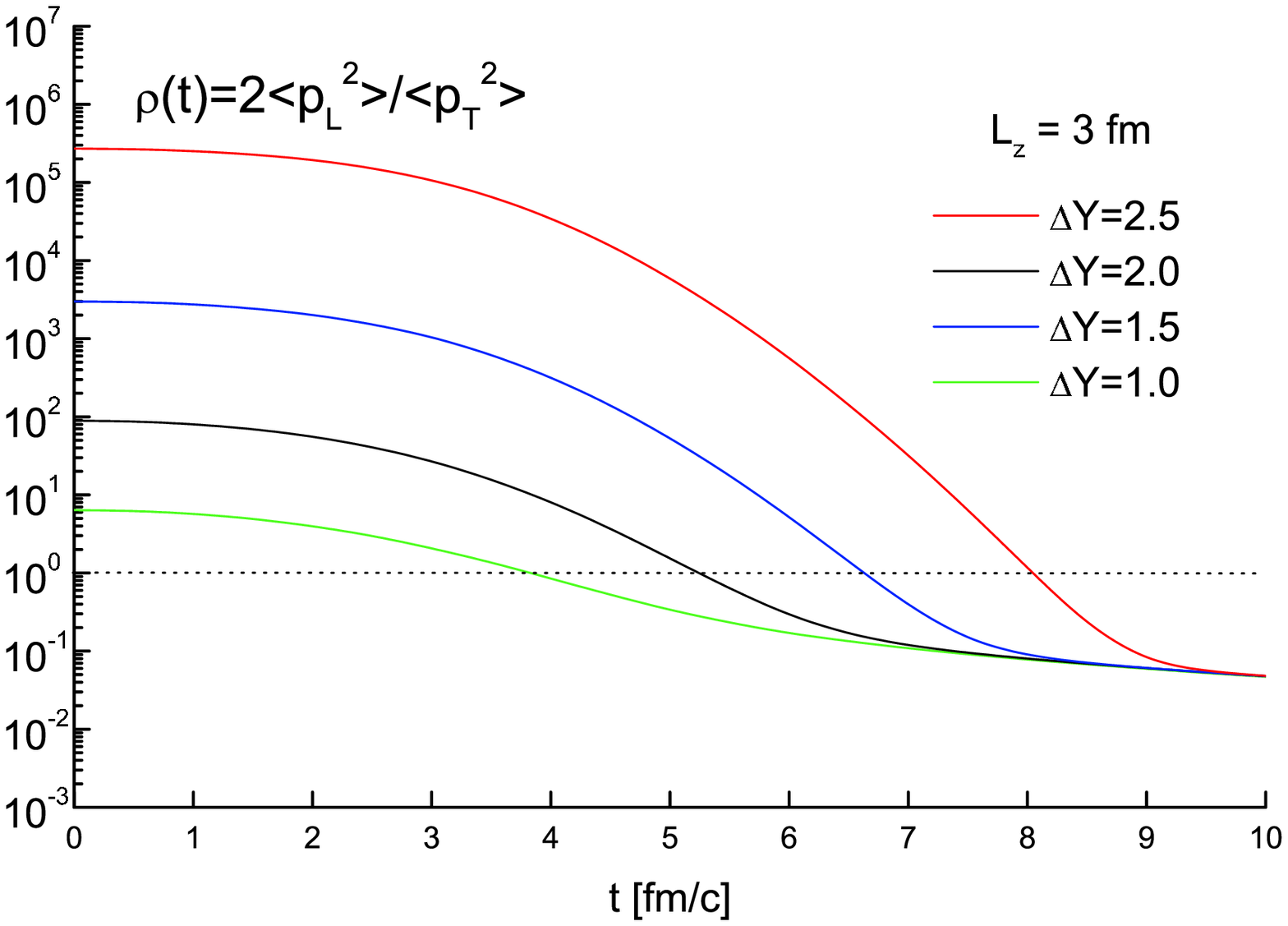}
\caption{(Color online) The momentum anisotropy as a function 
of time for four values of the rapidity distribution width 
$\Delta\!Y$. The most upper line corresponds to $\Delta\!Y=2.5$,
the lower one to $\Delta\!Y=2.0$, etc. The longitudinal size of 
the box $L_z$ equals 3 fm.
\label{fig-asym-p2}} 
\end{minipage}
\end{figure}

The momentum distribution of particles in the box is 
characterized by the parameter 
\be
\label{eccen-p-def}
\rho(t) = \frac{2\langle p_z^2 \rangle}
{\langle p_T^2\rangle } \;,
\ee
where 
$$
\langle p_z^2 \rangle = \frac{1}{n} \int d^3r {\cal O}({\bf r})
\frac{d^3p}{(2\pi)^3}\:p_z^2 f(t,{\bf r},{\bf p}) \;, \;\;\;\;
n = \int d^3r {\cal O}({\bf r})
\frac{d^3p}{(2\pi)^3}\: f(t,{\bf r},{\bf p}) \;,
$$
with analogous formula for $\langle p_T^2\rangle$; $f(t,{\bf r},{\bf p})$ 
is given by Eq.~(\ref{dis-fun}). For the prolate distribution, one has 
$\rho > 1$, for the isotropic one, there is $\rho = 1$, and finally, for 
the oblate distribution, we have $\rho < 1$. Since the parameter $\rho$ 
is sensitive only to the second moments of the momentum distribution, 
$\rho = 1$ is a necessary but not a sufficient condition for 
an isotropy of the distribution. 

The calculations simplify when the system is cylindrically symmetric
in coordinate space that is $\sigma_x = \sigma_y = \sigma_T$ and 
$L_x = L_y = L_T$. Then, the anisotropy parameter (\ref{eccen-p-def}) 
is given by
\be
\label{eccen-p}
\rho(t) = \frac{\int dY \,{\rm sh}^2Y \: G(t,Y)}{\int dY G(t,Y)} \;,
\ee
where
$$
G(t,Y) \equiv \exp \bigg[- 6\Big(\frac{1}{L_T^2 + 12\sigma_T^2} 
\frac{1}{{\rm ch}^2 Y}
+ \frac{1}{L_z^2 + 12\sigma_z^2} {\rm th}^2Y \Big) t^2 
- \frac{Y^2}{2 \Delta\!Y^2 }\bigg] \;.
$$
One easily computes the initial value of $\rho$ as 
$\rho (0) = e^{2\Delta\!Y^2}-1$. It is also of interest to see how 
the energy density $e(t)$ in the box decreases when the momentum 
anisotropy evolves. A simple calculation provides
$$
\frac{e(t)}{e(0)} = \frac{e^{-\Delta\!Y^2/2}}{\sqrt{2\pi}\Delta\!Y}
\int dY \, {\rm ch}Y \, G(t,Y) \;.
$$

\begin{figure}[t]
\begin{minipage}{83mm}
\includegraphics*[width=83mm]{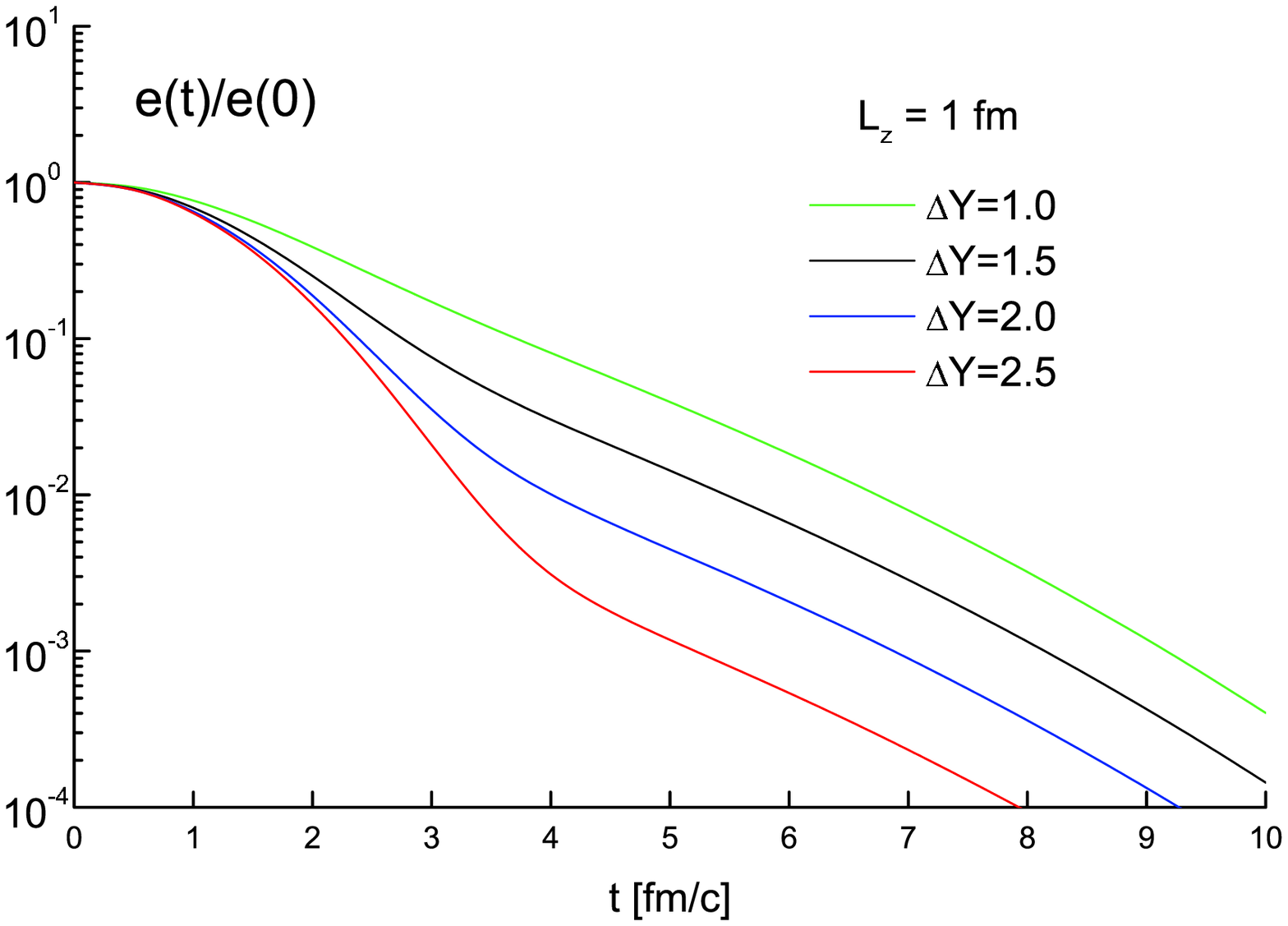}
\caption{(Color online) The relative energy density as a function 
of time for four values of the rapidity distribution width 
$\Delta\!Y$. The most upper line corresponds to $\Delta\!Y=1.0$,
the lower one to $\Delta\!Y=1.5$, etc. The longitudinal size of 
the box $L_z$ equals 1 fm.
\label{fig-en-rel1}} 
\end{minipage}\hspace{4mm}
\begin{minipage}{83mm}
\includegraphics[width=83mm]{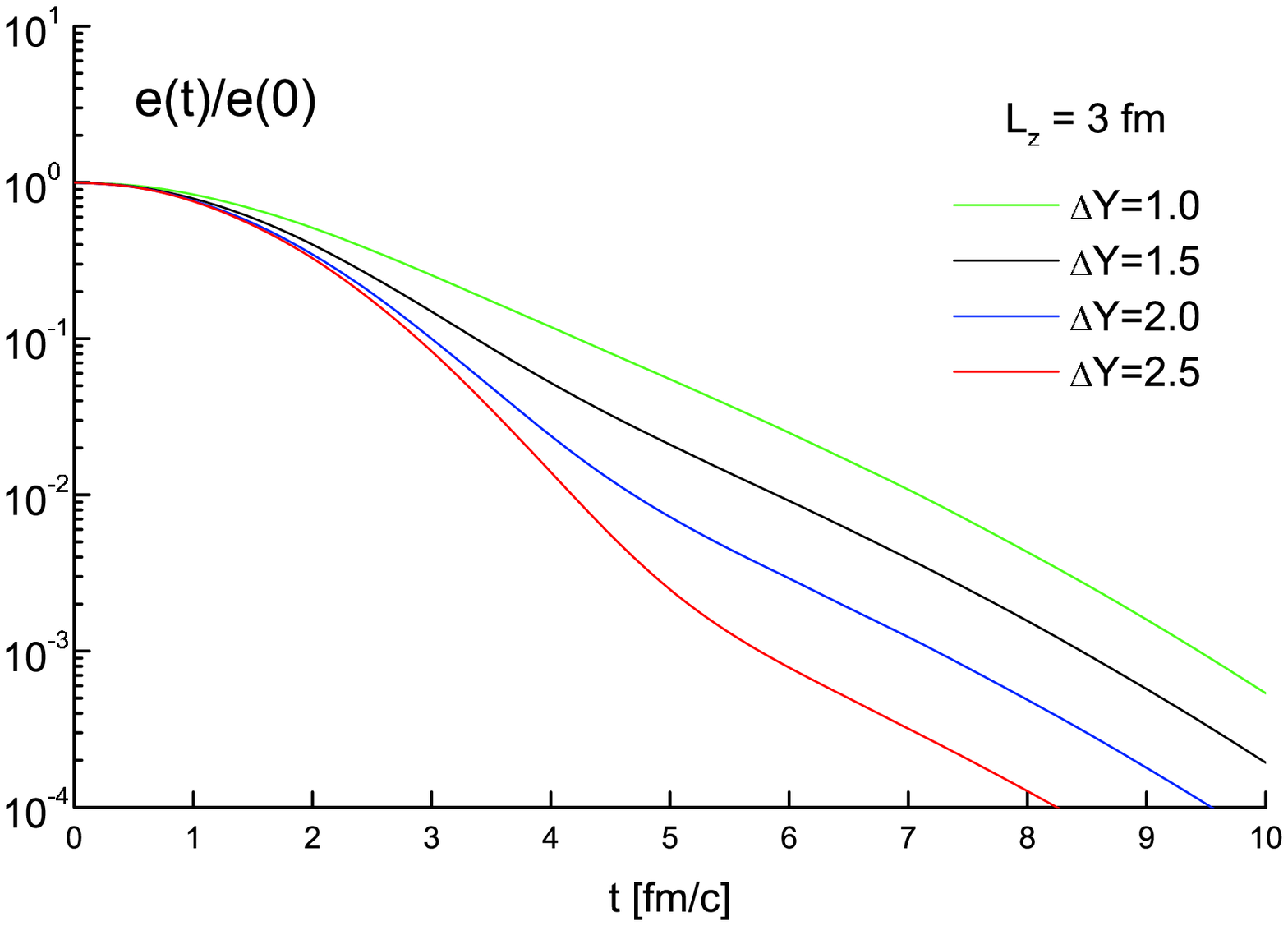}
\caption{(Color online) The relative energy density as a function 
of time for four values of the rapidity distribution width 
$\Delta\!Y$. The most upper line corresponds to $\Delta\!Y=1.0$,
the lower one to $\Delta\!Y=1.5$, etc. The longitudinal
size of the box $L_z$ equals 3 fm.
\label{fig-en-rel2}} 
\end{minipage}
\end{figure}

To compute $\rho(t)$ and $e(t)/e(0)$, the parameters $\sigma_T$, 
$\sigma_z$, $L_T$ and $L_z$ have to be chosen. As well known, the 
nuclear density of heavy nuclei is well described by the Woods-Saxon 
formula which can be roughly approximated by the sharp-sphere 
parameterization with the radius $R$ which for the heaviest nuclei 
equals about 7 fm. The mean square radius for the sharp-sphere 
parameterization equals $\langle {\bf r}^2 \rangle = 3R^2/5$. 
Therefore, we choose the widths of the Gaussian distribution 
$\sigma_T=\sigma_x=\sigma_y$ to be equal to 
$R/\sqrt{5} \approx 3 {\rm fm}$. The transverse size of the box
$L_T$ is assumed to coincide with $\sigma_T$. The longitudinal
width of the interaction zone $\sigma_z$ is chosen as 1 fm.
The calculations of $\rho(t)$ and $e(t)/e(0)$ are performed
for $L_z = 1\;{\rm fm}$ and $L_z = 3\;{\rm fm}$. The results
are shown in Figs.~\ref{fig-asym-p1}, \ref{fig-asym-p2}, 
\ref{fig-en-rel1} and \ref{fig-en-rel2}.

As already mentioned, the width of rapidity distribution 
of charged pions produced in Au-Au collisions at 
$\sqrt{s} = 200$ GeV is $\Delta\!Y =2.3$ \cite{Bearden:2004yx}.
One expects that the rapidity of produced partons is even 
broader. Therefore, the results shown in Figs.~\ref{fig-asym-p1}, 
\ref{fig-asym-p2} for $\Delta\!Y =2.5$ seem to be relevant
for heavy-ion collisions at RHIC. In such a case, it takes 
6-8 fm/$c$ to have an oblate momentum distribution. As
Figs.~\ref{fig-en-rel1}, \ref{fig-en-rel2} show, the energy 
density in the box is then decreased by a large factor - the 
system is much diluted.

One argues that particles, which are produced with rapidity $Y$, 
materialize only at a finite proper time $\tau$ and space-time 
rapidity $\eta = Y$. Keeping in mind that
$$
\tau = \sqrt{t^2 -z^2} \;, \;\;\;\;\;
\eta = \frac{1}{2}{\rm ln}\frac{t+z}{t-z} \;,
$$
one finds 
$$
t= \tau \: {\rm ch}\eta \;, \;\;\;\;\;
z = \tau \: {\rm sh}\eta \;.
$$
When the formation time $\tau$ is 0.3 fm/$c$ and 
$\eta = Y = 2.5$, one obtains $t = 1.8\;{\rm fm}/c$ and 
$z = 1.8\;{\rm fm}$. Thus, partons with $Y = 2.5$ materialize 
beyond the box of $L_z \le 1\;{\rm fm}$. To take into account
the effect of finite time formation in our analysis, we simply 
eliminate from the distribution function (\ref{dis-fun}), the 
partons of the proper time smaller than $\tau$. In other words, 
the function (\ref{dis-fun}) is multiplied by 
$\Theta (t\sqrt{1 - v_z^2}-\tau)$ and high-rapidity partons
are effectively excluded. 

\begin{figure}[t]
\begin{minipage}{83mm}
\includegraphics*[width=83mm]{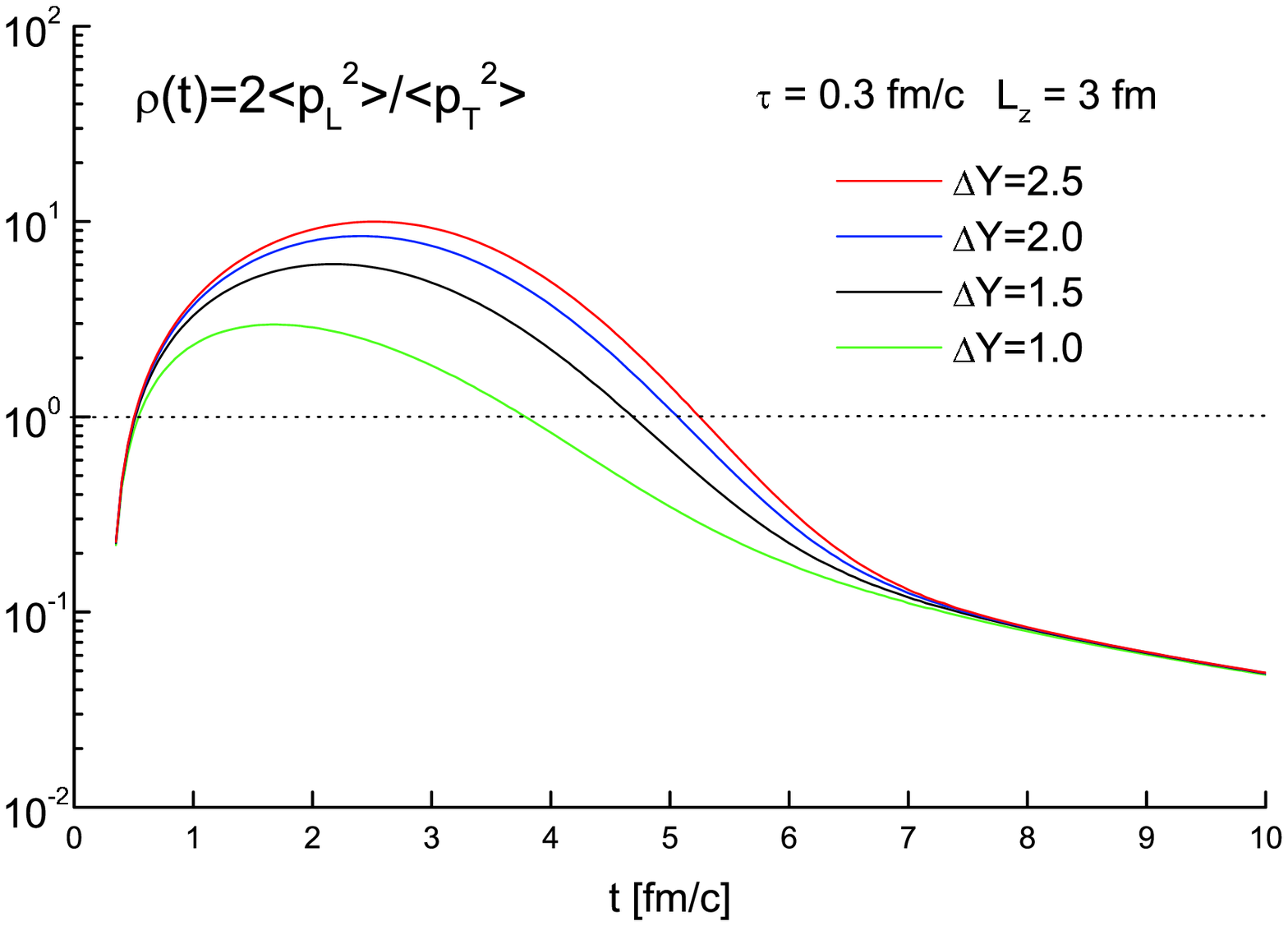}
\caption{(Color online) The momentum anisotropy as a function 
of time for the formation time $\tau = 0.3\;{\rm fm}/c$ and 
four values of the rapidity distribution width $\Delta\!Y$.
The most upper line corresponds to $\Delta\!Y=2.5$,
the lower one to $\Delta\!Y=2.0$, etc. The longitudinal
size of the box $L_z$ equals 3 fm.
\label{fig-asym-p-for3}} 
\end{minipage}\hspace{4mm}
\begin{minipage}{83mm}
\includegraphics[width=83mm]{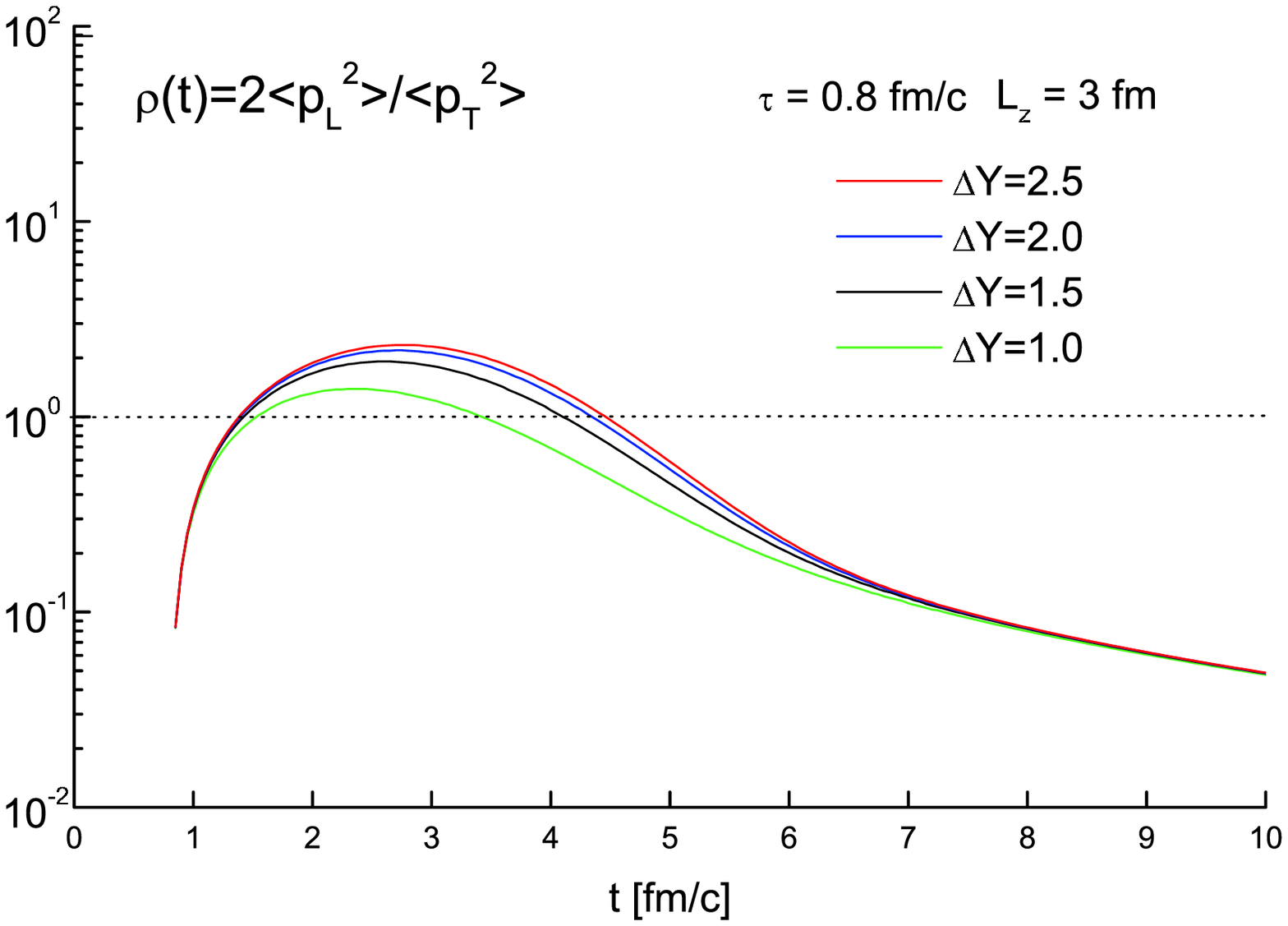}
\caption{(Color online) The momentum anisotropy as a function 
of time for the formation time $\tau = 0.8\;{\rm fm}/c$ and 
four values of the rapidity distribution width $\Delta\!Y$.
The most upper line corresponds to $\Delta\!Y=2.5$,
the lower one to $\Delta\!Y=2.0$, etc. The longitudinal
size of the box $L_z$ equals 3 fm.
\label{fig-asym-p-for8}} 
\end{minipage}
\end{figure}

The temporal evolution of the momentum asymmetry $\rho$, 
which takes into account the finite formation time, is shown  
in Figs.~\ref{fig-asym-p-for3} and \ref{fig-asym-p-for8} for
$\tau = 0.3\;{\rm fm}/c$ and $\tau = 0.8\;{\rm fm}/c$,
respectively. As previously, $\sigma_T = L_T = 3\;{\rm fm}$,
$\sigma_z = 1\;{\rm fm}$ and $L_z = 3\;{\rm fm}$. As seen,
the effect of finite formation time is very significant.
At the beginning the momentum distribution is oblate 
as partons with small $Y$ appear in the box earlier than
those with larger $Y$. After some time the momentum distribution 
is prolate, and it becomes again oblate due to the system's 
expansion only after 3-5 fm/$c$.

\section{Conclusions and Discussion}


As discussed in Sec.~\ref{sec-x-asym}, the parton system produced 
in nucleus-nucleus collisions has to be equilibrated before 1.5 fm/$c$.
Otherwise the eccentricity is too much reduced and the ideal 
hydrodynamics significantly underestimates experimental data of the 
elliptic flow. The results from Sec.~\ref{sec-p-asym} show that at 
the time 1.5 fm/$c$ the local momentum distribution is still prolate. 
Therefore, we conclude that just before local equilibrium is reached, 
the parton momentum distribution is elongated along the beam. Let us 
consider how reliable is the conclusion. 

First of all, if our rather conservative estimate of the upper limit 
of equilibration time of $t_{\rm eq}= 1.5 \;{\rm fm}/c$ is replaced 
by more elaborated estimate of 0.6 fm/$c$ given in \cite{Heinz:2004pj}, 
our conclusion is really safe - it seems impossible to build up the 
oblate momentum distribution in such a short time. 

One wonders whether the formation time $\tau$ can be extended. 
It should remembered, however, that in the case of finite $\tau$, the 
initial coordinate-space eccentricity does not occur at $t=0$ but at 
$t = \tau$ (for mid rapidity particles). And the interval of time 
when the system reaches equilibrium is reduced to $t_{\rm eq} - \tau$. 
So, the longer $\tau$, the fast thermalization is more and more
difficult to understand.

Temporal evolution of the momentum anisotropy is faster when $\sigma_z$, 
which is the initial longitudinal localization of produced partons, is 
reduced. However, for $L_z= 3 \;{\rm fm}$, the time when $\rho = 1$ is 
shorter only by 25\% when $\sigma_z$ decreases from 1 fm to 0.5 fm. So, 
our conclusion remains unchanged. We note that $\sigma_z$ should not be 
confused with the longitudinal localization of valence quarks of incoming 
nuclei which, due to the Lorentz contraction, is $\sigma_T/\gamma$ with 
$\gamma$ being the Lorentz factor. Our $\sigma_z$ corresponds to the 
{\em produced} partons. Therefore, it cannot be too small, as the wee 
partons, which are localized beyond the contracted volume of incoming 
nuclei, effectively participate in nucleus-nucleus collisions.  

Our free-streaming model of massless partons is obviously very 
naive. Partons, which are produced at the collision early stage, 
often carry a large virtuality acting as a mass. If the parton mass 
is taken into account both the coordinate and momentum space evolutions 
are slowed down. Inter-parton interactions presumably lead to a similar 
effect. However, we cannot see a good reason that the coordinate space 
evolution is slowed down much more than the momentum space evolution. 
Therefore, our conclusion seems to be rather safe.

Obviously it is desirable to improve our free-streaming model
but the problem is rather difficult. Color Glass Condensate 
(CGC) approach (for a review see \cite{Iancu:2003xm}), which 
is the best developed effective theory to study the early stage 
of heavy-ion collision, is not well suited for the problem.
The valence quarks are treated in CGC as classical
passive sources of small $x$, highly populated gluons which 
are described in terms of classical fields. Partons with
sizeable $x$, which crucially influence the momentum 
distribution, are essentially absent in CGC.

Numerous studies of equilibration of parton systems with the 
initially oblate momentum distribution, which are reviewed in 
\cite{Mrowczynski:2005ki}, are theoretically well founded as
the oblate system is diluted and decohered. However, according 
to our conclusion, these studies do not actually explain how 
local equilibrium is reached in heavy-ion collisions but rather 
how the equilibrium is sustained. We believe that the plasma
equilibration, in particular the color instabilities, which are 
supposed to speed up the process of thermalization, should be 
analyzed as in \cite{Mrowczynski:1994xv,Randrup:2003cw} that
is in the systems with prolate momentum distribution. We are 
aware that kinetic theory is hardly applicable to such dense, 
inhomogeneous and partially coherent systems but Nature does 
not seem to be bothered by our theoretical difficulties. 

\begin{acknowledgments}

We are grateful to Uli Heinz for helpful correspondence. This 
work was partially supported by Polish Ministry of Science and
Higher Education under grant 1 P03B 127 30.

\end{acknowledgments}


\end{document}